\documentclass[12pt]{iopart}
\usepackage{xspace,amsfonts,amsthm,amssymb,amsbsy}
\usepackage{graphicx}
\usepackage{amssymb}
\usepackage{color}

\expandafter\let\csname equation*\endcsname\relax
\expandafter\let\csname endequation*\endcsname\relax
\usepackage{amsmath}

\usepackage[usenames,dvipsnames]{xcolor}
\usepackage[colorlinks=true,citecolor=Cerulean,linkcolor=RubineRed,urlcolor=Cerulean]{hyperref}
\usepackage{iopams}

\definecolor{green}{rgb}{0.2,0.7,0.4}
\newcommand{\ket}[1] {\left\vert #1 \right\rangle}
\newcommand{\bra}[1] {\left\langle #1 \right|}
\newcommand{\braket}[2] {\langle #1 | #2 \rangle}

\newcommand{\dg}{^{\dagger}}

\newcommand{\vac}{\ket{v}}

\newcommand{\bP}{\mathbf{P}}

\newcommand{\bQ}{\mathbf{Q}}
\newcommand{\bR}{\mathbf{R}}

\newcommand{\fB}{\mathcal{B}}
\newcommand{\fD}{\mathcal{D}}

\graphicspath{{../../images/}}

\begin{document}

\title{Composite boson signature in the interference pattern  of  atomic dimer condensates}

\author{Shiue-Yuan Shiau$^1$, Aur\'elia Chenu$^{2-3}$\footnote{Current address: Ikerbasque Research Fellow at the Donostia International Physics Center,  E-20018 San Sebasti\'an, Spain} and Monique Combescot$^4$}

\address{$^1$ Physics Division, National Center for Theoretical Sciences, Hsinchu 30013, Taiwan}
\address{$^2$ Theory Division, Los Alamos National Laboratory, MS-B213, Los Alamos, NM 87545, USA}
\address{$^3$ Department of Chemistry, Massachusetts Institute of Technology,  77 Massachusetts Avenue, Cambridge, MA 02139, USA}
\address{$^4$ Sorbonne Universit\'e, Institut des NanoSciences de Paris,  CNRS, 4 place Jussieu, 75005 Paris}

\ead{shiau.sean@gmail.com}



\begin{abstract}
We predict the existence of high frequency modes in the interference pattern of two  condensates made of fermionic-atom dimers. These modes, which result from fermion exchanges between condensates,  constitute a striking signature of the dimer composite nature. From the 2-coboson spatial correlation function, that we derive analytically, and the Shiva diagrams that visualize many-body effects specific to composite bosons, we identify the physical origin of these high frequency modes and determine the conditions to see them experimentally by using bound fermionic-atom pairs trapped on optical lattice sites. The dimer granularity which appears in these modes comes from Pauli blocking that  prevents two dimers to be located at the same lattice site.

\end{abstract}
\submitto{\NJP}

\maketitle

\section{Introduction\label{sec:X1}}

All particles consisting of an even number of fermions are  boson-like. Although this property merely derives from a mathematical fact---the  particle creation operators commute---the boson-like nature of the particles bears a strong consequence: they must undergo Bose-Einstein condensation (BEC). This physical effect has been  observed for dilute gases of ultracold bosonic atoms~\cite{Anderson1995sci,DavisPRL1995,BradleyPRL1995,Zwierlein2005Nat}, decades after Einstein predicted it. 
An oscillatory behavior showing the condensate coherence  has also been observed in the interference pattern of two  condensates of bosonic atoms \cite{Wang2005,Andrew1997,Vogels2003,Shin2005}.\

The fact that boson-like particles are made of fermions shows up in nontrivial ways. This  composite nature mathematically appears through the commutator of their destruction and creation operators, $\big[\fB_i,\fB_j\dg\big]_-=\delta_{i,j}-D_{i,j}$. While   this commutator reduces to the delta term for elementary bosons, the $D_{i,j}$ operator is responsible for fermion exchanges with other particles~\cite{MoniqPhysreport,MoniqueSeanbook}.  Although these exchanges are commonly neglected, the  composite bosons (cobosons) then reducing to point-like structureless elementary bosons, they can have significant consequences.  
The study of the particle composite  nature in the field of semiconductor excitons was pioneered by Keldysh and Kozlov \cite{Keldysh1968} as early as 1968.

In a gas, the coboson centers of mass are delocalized over the system volume $L^D$, where $D$ is the space dimension, while fermion exchanges occur over the coboson volume $a_B^D$, where $a_B$ is its Bohr radius. So, many-body effects induced by  fermion exchanges  
 between $N$ cobosons are controlled by the dimensionless parameter
 \begin{equation}
    \eta= N \Big(\frac{a_B}{L}\Big)^D\, .\label{1}
     \end{equation}
This  leads us to conclude that the particle composite nature can only have significant consequences  for a dense gas  at the scale of the coboson size \footnote{Note that this composite nature also appears at the two-body level, through a significant reduction of the dimer-dimer scattering length by the repeated effective coboson-coboson interaction in which fermion exchange plays a key role~\cite{Petrov2004}.}. 
Sizeable $\eta$'s are difficult to reach for cold atoms due to the very small atom size; yet a $^6$Li$_2$ dimer condensate with density reaching $0.3$ on the dimer-dimer scattering length  scale has been reported~\cite{Bourdel2004}.

 By contrast, in the case of semiconductor Wannier excitons, with size two orders of magnitude larger than  typical atoms, values of $\eta$ as large as 1, or even larger, are easy to reach---although for such large $\eta$'s, excitons  dissociate into an electron-hole plasma. The Wannier exciton composite nature has been shown to have a significant impact on the physics of excited semiconductors.  Among its noticeable consequences, we can cite the exciton optical Stark effect~\cite{Mysyrowicz,Hulin,MCPR} and the coexistence of dark and bright condensates~\cite{alloing,dubin2017} that results~\cite{Monique2007} from the coupling, through fermion exchange, between bright excitons with spin $\pm 1$ and dark excitons with spin $\pm 2$. \
  
  It has been shown from the study of Wannier excitons, Frenkel excitons and Cooper pairs~\cite{MoniqueSeanbook}, that the dimensionless parameter which rules  composite boson many-body effects physically corresponds to $\eta = N / N_{\rm max}$, where $N_{\rm max}$ is the number of cobosons that the sample can accommodate. For Wannier excitons, this number is $(L/a_B)^3$, because for a higher number,  excitons overlap and dissociate into an electron-hole plasma. In the case of Frenkel excitons, made of on-site excitations in a periodic lattice, $N_{\rm max}$ is the number $N_s$ of lattice sites in the sample at hand. 

 As a result, a more controllable platform to get sizable $\eta$ is not to use a gas but an optical lattice, as previously considered to study Hong-Ou-Mandel-like interferences~\cite{Tichyprl2012}. Indeed, dense samples in which each lattice potential well  traps a single dimer have already been made, with $\eta\sim 0.4$ for $^{40}$K$_2$ fermionic-atom dimers~\cite{Thilo2006}, and $\eta\sim 0.5$  or $0.8$ for $^{87}$Rb$_2$~\cite{Volz2006} or $^{133}$Cs$_2$~\cite{Danzl2010} bosonic-atom dimers. In the case of heteronuclear dimers, dense optical lattice samples of RbCs \cite{Lukas2017} and KRb \cite{Moses2,Moses}  have also been reported. These studies open an exciting route in the field of cold atoms, toward studying the  rich yet essentially unknown world of many-body effects resulting from dimensionless fermion (or boson) exchanges, that is, exchanges occurring between quantum particles in the absence of energy-like particle-particle interaction. 

Motivated by the pioneering studies of  condensate coherence properties in the case of bosonic atoms \cite{Wang2005,Andrew1997,Vogels2003,Shin2005}, we here investigate the effect of the particle composite nature on the interference pattern of two  condensates made of fermionic-atom dimers. We first give arguments to find their signature in the spatial correlation function; next, we provide hints on how this function can be analytically derived, and finally we discuss the relevant limits.

To do it, we consider  $N$ pairs  of  fermionic atoms, $\alpha$ and $\beta$, in different hyperfine states, these atoms being  trapped in an optical lattice having $N_s$ sites. Their Hamiltonian reads as \cite{Tichyprl2012}
\begin{eqnarray}
H&=&H_0-U_{\alpha\beta}\sum_{i=1}^{N_s} a\dg_{\bR_i}b\dg_{\bR_i}b_{\bR_i}a_{\bR_i}+\sum_i\sum_{j\not= i}V_{\bR_{j}-\bR_{i}}a\dg_{\bR_j}b\dg_{\bR_j}b_{\bR_i}a_{\bR_i}\,,\label{Hamil10}\\
H_0&=&\sum_{i=1}^{N_s}\varepsilon_\alpha a\dg_{\bR_i}a_{\bR_i}+\sum_{i=1}^{N_s}\varepsilon_\beta b\dg_{\bR_i}b_{\bR_i}\,.
\end{eqnarray}
The $H_0$ eigenstates for one fermionic-atom pair are $a\dg_{\bR_i}b\dg_{\bR_j}\vac$ with energy $\varepsilon_\alpha+\varepsilon_\beta$, where $\vac$ denotes the vacuum. 
The energy levels, $(\varepsilon_\alpha,\varepsilon_\beta)$, depend on the optical lattice potential. The strength  of the atom-atom attractive potential, $U_{\alpha\beta}>0$, can be varied  through Feshbach resonance, allowing a control on the spatial extension $a_B$ of a  bound pair. 
For large $U_{\alpha\beta}$  attraction, the lowest-energy subspace reduces to the $N$ states, $a\dg_{\bR_i}b\dg_{\bR_i}\vac$, with energy $\varepsilon_\alpha+\varepsilon_\beta-U_{\alpha\beta}$, each lattice site $\bR_i$ possibly hosting one  bound pair, with creation operator  $\fB\dg_{\bR_i} = a\dg_{\bR_i} b\dg_{\bR_i} $.  
These  pairs are structureless when the lattice period, which is equal to half the optical laser  wavelength $\lambda$, is  large compared to the relative-motion extension $a_B$ of a  bound pair. This inequality, $\lambda\gg a_B$, can be easily fulfilled in optical lattice experiments, as in the case of  $^{40}$K$_2$ dimers~\cite{Thilo2006}.

The third term of Eq.~(\ref{Hamil10}) describes inter-site interaction. Its $V_{\bR_{j}-\bR_{i}}$ strength  can be changed by changing the laser intensity which is proportional to the lattice  potential depth $V_0$. Due to this inter-site interaction,  bound fermionic-atom pairs delocalize over the whole lattice. To justify our  consideration of  one bound pair per lattice site at most,  the inter-site interaction has to be   small compared to the  on-site  repulsion. This condition can be achieved  when  the lattice  potential depth $V_0$ is large compared to  the recoil energy $E_R=\hbar^2k^2/2m_a$, with $k=2\pi/\lambda$ and $m_a$  the atom mass, as previously shown in the case of elementary bosons~\cite{Jaksch98}. 


  In the following, we shall refer to  fermionic atoms simply as fermions, and refer to  delocalized bound pairs as dimers. The dimer ground-state creation operator reads 
\begin{equation}
 \fB\dg_{\bQ} = \sum_{i=1}^{N_s} \fB\dg_{\bR_i} \braket{\bR_i}{\bQ}
 \end{equation} with a distribution which is flat, $\braket{\bR_i}{\bQ} = e^{i\bQ\cdot \bR_i} /\sqrt{N_s}$, in the large lattice limit. The small dimer spatial extension $a_B$ being small, it  disappears from the dimensionless parameter that controls many-body effects. Instead, this dimensionless parameter for $N$ dimers in an optical lattice having $N_s$ sites reads, like for Frenkel excitons~\cite{MoniqueSeanbook,MoniquePRB2008,Agra2008}, as
 \begin{equation}
    \eta= \frac{N}{N_s} .\label{1'}
     \end{equation}
     
 To produce two condensates made from such dimers,  we propose to first prepare a single condensate  in an optical lattice having a lattice spacing  equal to a few hundreds of nanometers, depending on the laser wavelength. Then, we ramp up a potential barrier in the middle of the lattice using an external field in order to split this condensate  into two spatially separated condensates, which are ultimately let to interfere, and we  measure the correlation function. The potential barrier in the middle of the experimental setup can be produced by imposing another optical lattice with a much larger lattice spacing. Lattice spacing of the order of a hundred micrometers has been experimentally produced~\cite{Fallani} by  using two  near-infrared laser beams with $\lambda$ wavelength  intersecting at a small angle $\theta$,  the lattice spacing being equal to $\lambda/2\sin(\theta/2)$. For nearly collimated beams, that is,  for $\theta\sim 0$, the lattice spacing  can become very large, of the order of hundreds  of micrometers.

We predict that, compared to elementary bosons, the interference pattern  has additional high frequency modes that come from fermion exchanges between condensates. These interferences constitute a striking signature of the dimer composite nature.   As these additional modes are many-body in essence, a sizeable $\eta$ is required to observe them. This is why previous experiments performed with two rather dilute condensates made of small bosonic atoms like rubidium \cite{Wang2005} or sodium \cite{Andrew1997,Vogels2003,Shin2005}, with momenta  $\bQ$ and $\bQ'=-\bQ$, have only seen interferences ruled by the momentum difference,  $\bQ-\bQ'=2\bQ$. Such interferences  can be obtained by taking the bosonic atoms as  elementary bosons \cite{Javanainen1996a,Javanainen1997a,Castin1997a}. 
The higher frequency modes we predict come from  fermion exchanges involving more than one dimer from each of the two $\bQ$ and $\bQ'$ condensates. They  can produce momentum differences $m(\bQ-\bQ')$ with $m\geq 2$: the  $m=2$ mode appears when  at least one fermion exchange in each condensate enters into play; so,  its amplitude is $\eta^2$ smaller than the $m=1$ mode. Fermion exchanges also affect the $m=(0,1)$ modes present for elementary bosons, but only through $\eta$ corrections in their prefactors.

\section{Physics of the problem}
 
 In this section, we use simple physical arguments to understand the form of the interference pattern resulting from the collision of two dimer condensates.  An appropriate way to derive this interference pattern is through the dimer-dimer spatial correlation function
  \begin{eqnarray}
\mathcal{G}^{(2)}_{N,N'}(\bR_1,\bR_2)=
\frac{\bra{\psi_{N,N'}} \fB\dg_{\bR_1} \fB\dg_{\bR_2} \fB_{\bR_2}  \fB_{\bR_1} \ket{\psi_{N,N'}}}{\langle \psi_{N,N'}|\psi_{N,N'}\,  \rangle}\,,\label{3}
\end{eqnarray}
the  two-condensate state made of  $N$ dimers of momentum $\bQ$ and $N'$ dimers of momentum $\bQ'$ reading as
 \begin{equation}
 \ket{\psi_{N,N'}}= (\fB\dg_{\bQ})^N (\fB\dg_{\bQ'})^{N'}\vac\,.
  \end{equation} 
 Taking $N'\neq N$ and $\bQ'\neq-\bQ$ makes the physics  easier to grasp.
  
  \begin{figure}[t!]
\centering
\includegraphics[trim=4.3cm 5.2cm 4cm 6.5cm,clip,width=4in]{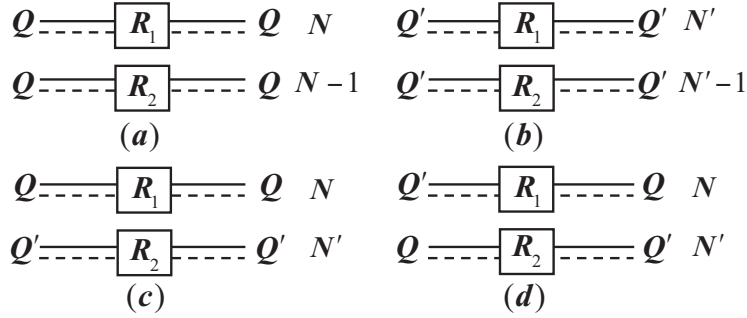}
\caption{\small Correlation function $ \mathcal{G}^{(2)}_{N,N'}(\bR_1,\bR_2)$ defined in Eq.~(\ref{3}), in the absence of fermion exchanges. A dimer is destroyed and recreated from the same fermion pair at the same position with same momentum (a,b,c) or with  different momenta (d), the total momentum being conserved. The oscillatory  $\cos (\bQ-\bQ')\cdot \bR_{12}$ term of Eq.~(\ref{4}) comes from (d). In Shiva diagrams, a coboson dimer is represented by a double line, the solid and dashed lines representing its two different fermionic atoms. See \cite{MicPRL2010} for a detailed description of Shiva diagrams.}
   \label{fig:1}
\end{figure}

Let us first deal without fermion exchanges, which corresponds to taking the dimers as elementary bosons. The  operator $\fB_{\bR_1}$ destroys one of the $(N+N')$ dimers of the $ \ket{\psi_{N,N'}}$ state at the $\bR_1$ site. Let this destroyed dimer have a momentum $\bQ$.  If $\fB_{\bR_2}$  also destroys a $\bQ$ dimer, these two detections generate a $N(N-1)$ factor from the number of ways to choose the two $\bQ$ dimers among $N$. Due to momentum conservation in the $\mathcal{G}^{(2)}_{N,N'}(\bR_1,\bR_2)$ numerator, $\fB\dg_{\bR_1}$ and $ \fB\dg_{\bR_2} $ must recreate two $\bQ$ dimers. The associated phase factors $e^{i \bQ \cdot \bR_1}$ and $e^{-i \bQ \cdot \bR_1}$ to detect a $\bQ$ dimer at $\bR_1$ then cancel; same at $ \bR_2$. So, the term in $N(N-1)$ does not bring any oscillatory contribution to $\mathcal{G}^{(2)}_{N,N'}(\bR_1,\bR_2)$ (Fig.~\ref{fig:1}a). In the same way, no oscillation occurs if two $\bQ'$ dimers (Fig.~\ref{fig:1}b) or one $\bQ$ dimer and one $\bQ'$ dimer (Fig.~\ref{fig:1}c) are destroyed and recreated at the same site, their counting factor being, respectively, $N'(N'-1)$ and $2NN'$, the $2$ coming from detecting the $\bQ$ dimer at $\bR_1$ or at $\bR_2$. So, we end up with a contribution to $\mathcal{G}^{(2)}_{N,N'}(\bR_1,\bR_2)$ equal to 
 \begin{equation}
 N(N-1)+N'(N'-1)+2NN'.
  \end{equation}

 For an oscillatory term to appear in the correlation function, the dimers destroyed and recreated at the $\bR_i$ site must have different momenta (Fig.~\ref{fig:1}d). The term in which $\fB_{\bR_1}$ destroys a $\bQ$ dimer and $\fB\dg_{\bR_1}$ recreates a $\bQ'$ dimer brings a factor $e^{i  \bR_1 \cdot( \bQ-\bQ')}$ with a $NN'$ prefactor coming from the number of ways to choose these $(\bQ, \bQ')$ dimers. To conserve momentum, $\fB_{\bR_2}$ then has to destroy a $\bQ'$ dimer and $\fB\dg_{\bR_2}$ to create a $\bQ$ dimer, which  brings a factor $e^{i  \bR_2 \cdot( \bQ'-\bQ)}$. So, we end up with a contribution to $\mathcal{G}^{(2)}_{N,N'}(\bR_1,\bR_2)$ equal to 
  \begin{equation}
  NN'\Big(e^{i ( \bR_1-\bR_2) \cdot( \bQ-\bQ')}+e^{i ( \bR_1-\bR_2) \cdot( \bQ'-\bQ)}\Big)=2NN'  \cos ( \bQ-\bQ') \cdot \bR_{12}\,,  \label{4}
   \end{equation}
where $ \bR_{12}= \bR_{1}- \bR_{2}$. This leads to the $m=1$ mode previously found for elementary bosons. Indeed, when $N=N'$  and $\bQ'=-\bQ$, the spatial correlation function for  elementary bosons corresponds to 
\begin{equation} 
\mathcal{\bar  G}^{(2)}_{N,N}(\bR_1,\bR_2)= \frac{(2N)!}{N_s^2(2N-2)!}\Big\{1+\frac{N}{2N-1} \cos 2\bQ\cdot \bR_{12}  \Big\}\,, \label{eq:G2_elementary}
\end{equation}
 as first given in Ref.~\cite{Javanainen1996a}. For completeness, we rederive in \ref{app:sec1} the $n$-particle correlation function for free elementary bosons using a different approach.
The amplitudes of the $m=(0,1)$ terms both scale as $N^2/N_s^2\sim \eta^2$ and no
other mode exists in the case of elementary bosons.\

\begin{figure}[t!]
\centering
\includegraphics[trim=0cm 6.5cm 0cm 5.5cm,clip,width=\columnwidth]{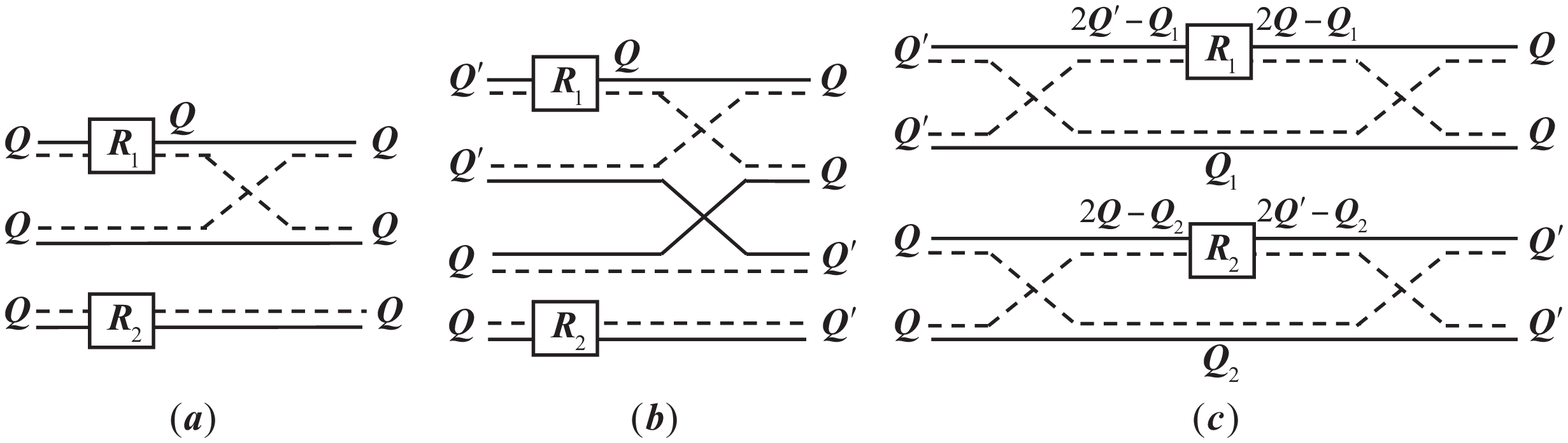}
   \caption{\small Shiva diagrams for fermion exchanges: (a) Exchanges within the $\bQ$ condensate do not change the momentum of the dimer detected at $\bR_1$; so, they do not lead to oscillatory terms. (b,c) Exchanges involving the two condensates can lead to dimers detected at the same site having momentum difference $m(\bQ'-\bQ)$, with  $m=1$  as in (b), or $m=2$ as in (c). Modes with $m\geq 2$ are the  signature of the dimer composite nature we predict.}
   \label{fig:2}
\end{figure}

   The dimer composite nature brings a far richer physics because $\fB_{\bR}$ can destroy any two  fermions. These two fermions can simply be the ones of the $\fB\dg_{\bQ}$ or $\fB\dg_{\bQ'}$ pair in $\ket{\psi_{N,N'}}$. Or, since identical fermions are indistinguishable, they can also be any pair resulting from fermion exchanges in the $\ket{\psi_{N,N'}}$ state. As exchanges conserve momentum, fermion exchange inside the $\bQ$ condensate does not change the momentum of the dimer detected at $\bR_1$ site,  as shown in the Shiva diagram of Fig.~\ref{fig:2}a. So, this does not bring any oscillatory contribution. However, as each fermion exchange brings a $1/N_s$ factor,  this term appears with a prefactor equal to
    \begin{eqnarray} 
   N(N-1)(N-2)/N_s\simeq \eta N(N-1)\,.
     \end{eqnarray}
 It thus is $\eta$ times smaller that the leading $N(N-1)$ term obtained
   in the absence of fermion exchange, shown in Fig.~\ref{fig:1}a.
    
By contrast, fermion exchanges between the $\bQ$ and $\bQ'$ condensates can lead to terms in which the detected dimers have a momentum difference equal to $m(\bQ-\bQ')$, with $m=1$ as in Fig.~\ref{fig:2}b, $m=2$ as in Fig.~\ref{fig:2}c, and $m\geqslant 3$ for exchanges involving more dimers from each condensate. The term corresponding to the Shiva diagram of Fig.~\ref{fig:2}b produces the same $ \cos (\bQ-\bQ')\cdot \bR_{12}$ oscillatory term as the one of Eq.~(\ref{4}), but its prefactor
  \begin{eqnarray} 
N(N-1)N'(N'-1)/N_s^2\simeq \eta \eta'NN'\,
  \end{eqnarray} 
 is $\eta \eta'$ times smaller due to the two exchanges it contains. More generally, the $m=1$ terms in which fermion exchanges enter  bring density-dependent corrections to the amplitude of the interference terms already present for elementary bosons.\

Momentum changes $m(\bQ-\bQ')$ with $m\geq 2$ are a definite signature of the condensate composite nature because they generate new oscillatory modes. The $2 (\bQ-\bQ')$ momentum change in Fig.~\ref{fig:2}c produces a $\cos 2(\bQ-\bQ')\cdot \bR_{12}$ contribution. This $m=2$ term has four exchanges, but two sums over $\bQ_1$ and $\bQ_2$ which cancel  two $N_s$ factors coming from exchanges; so, it also appears with a prefactor $N(N-1)N'(N'-1)/N_s^2\simeq \eta \eta' NN'$. Note that in order to produce these higher modes, similar exchanges must occur in  the $\bQ$ and the $\bQ'$ condensates; so, $m$ cannot be larger than $\textit{Min}\{N,N'\}$.

Last but not least, the fact that two identical fermions cannot be at the same site must hold sway over the possibility of seeing two dimers at the same site. Indeed, $(a\dg_{\bR})^2=0$ imposes $(\fB\dg_{\bR})^2=0$. So, $\mathcal{G}^{(2)}_{N,N'}(\bR_1,\bR_2)$ must cancel for $\bR_1= \bR_2$.  \

All this shows that fermion exchanges inside each condensate and between the two condensates change the interference pattern compared to that obtained with elementary bosons in three ways:

 (i) the amplitudes of the elementary-boson terms have many-body corrections reading in powers of densities, $\eta$ and $\eta'$; 
 
 (ii) higher oscillatory modes in $\cos m(\bQ-\bQ')\cdot \bR_{12}$ with $m=(2,3,\dots)$ up to $\textit{Min}\{N,N'\}$ appear, with ever weaker amplitudes;
 
  (iii) a dip at the scale of the optical lattice constant  exists for $\bR_1= \bR_2$.

 The spatial correlation function for fermionic-atom dimers  thus has to read 
\begin{equation} \label{eq:G2}
\mathcal{G}^{(2)}_{N,N'}(\bR_1,\bR_2)=\big(1-\delta_{\bR_1,\bR_2}\big)\!\!\sum_{m=0}^{\textit{Min}\{N,N'\}}A_{N,N'}^{(m)} \cos \Big (m(\bQ-\bQ')\cdot \bR_{12}\Big)\, .
\end{equation} 
 The amplitudes of the $m=(0,1)$ modes are equal to the elementary-boson values within density corrections coming from fermion exchanges, namely
 $A_{N,N'}^{(0)}\simeq (\eta+\eta')^2(1+\mathcal{O} (\eta,\eta'))$ and $A_{N,N'}^{(1)}\simeq 2 \eta \eta'(1+\mathcal{O} (\eta\eta'))$. The larger number of exchanges required for the $m= 2$ mode appears in its amplitude which scales as $A_{N,N'}^{(2)}\sim  (\eta \eta')^2 $ within density corrections. And so on for larger $m$.

 \section{Theoretical approach}
 For dimers characterized by a single quantum index $\bQ$, as dimers in an optical lattice, it is possible to perform an \textit{exact} calculation of the $n$-coboson spatial correlation function, $\mathcal{G}^{(n)}_{N,N'}(\bR_1, \dots ,\bR_n)$, in spite of the quite tricky fermion exchanges that occur not only within each condensate, but also between the $\bQ$ and $\bQ'$ condensates---from which  the most interesting physics arises. To do it, we have developed an original procedure in which $ \ket{\psi_{N,N'}}$ is written in terms of the generalized coherent states $\ket{\phi_{z,z'}}=e^{z\fB\dg_\bQ+z' \fB\dg_{\bQ'}}\vac$ where $(z,z')$ are complex scalars, as explained in \ref{app:sec2}. Using it, we can obtain $\mathcal{G}^{(n)}_{N,N'}(\bR_1, \dots ,\bR_n)$ analytically for arbitrary $n$. As its   expression is extremely complicated, even for $n=2$,  we shall here only discuss two limiting cases that best illustrate the involved physics, and refer the interested readers to \ref{app:sec2} for the general form.\

  
 \section{Analytical results for limiting cases}
 
 For $n=1$, the function $\mathcal{G}^{(1)}_{N,N'}(\bR_1)$ stays equal to its elementary-boson value, $(N+N')/N_s$, which physically corresponds to the total dimer density of the two condensates at the scale of the lattice site number $N_s$.
This result follows from the fact that (i) $\mathcal{G}^{(1)}_{N,N'}(\bR_1)$ does not depend on $\bR_1$, and (ii) the mean value of the number operator $\sum_\bQ \fB\dg_\bQ \fB_\bQ =  \sum_\bR \fB\dg_\bR \fB_\bR$ in the $\ket{\psi_{N,N'}}$ state is equal to the total dimer number, $N+N'$. \

\begin{figure}[t!]
\centering
\includegraphics[trim=0cm 0cm 0cm 0cm,clip,width=0.7\columnwidth]{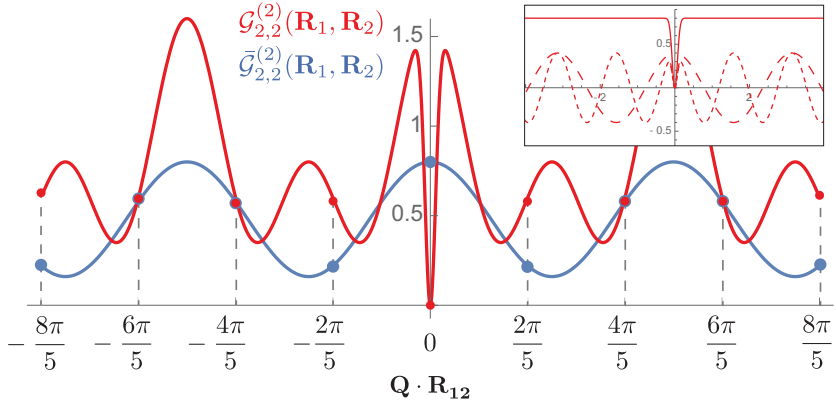}
   \caption{Spatial correlation function $\mathcal{G}_{2,2}^{(2)}(\bR_1, \bR_2)$ for  $(2,\bQ)$ and $(2,-\bQ)$   cobosons when $N_s=5$ [red circles, Eq.(\ref{G2_2})].  For periodic lattice, the red circles correspond to the set of values for $\bQ\cdot \bR_{12}$. The blue circles correspond to elementary bosons [Eq.(\ref{eq:G2_elementary})]. The curves are plotted by taking $\bQ\cdot \bR_{12}$  continuous to guide the eye.  The inset shows  each term in  Eq. (\ref{G2_2}).  }
   \label{fig:3}
\end{figure}

To grasp  how fermion exchanges affect the interference pattern of two  condensates, let us consider the simplest case in which the predicted $m=2$ oscillatory mode exists, that is $N=N'=2$. The explicit expression of the spatial correlation function then reads [see Eq.~(\ref{app:generalGNN})]
\begin{equation}\label{G2_2}
\mathcal{G}_{2,2}^{(2)}(\bR_1,\bR_2)=\frac{(1-\delta_{\bR_1,\bR_2})}{N_s^2}\Big\{12(1+x_0)   +8(1+x_1) \cos (\bQ-\bQ')\cdot \bR_{12}  +x_2  \cos 2(\bQ-\bQ')\cdot \bR_{12}\Big\}.  
\end{equation}
The changes from the elementary-boson result given in Eq.~(\ref{eq:G2_elementary}), induced by exchanges,  appear in  the $x_i$'s, as illustrated in Fig.~\ref{fig:3}. They  read $x_0=(5/3N_s-15/N_s^2+36/N_s^3)/F_{2,2}$, $x_1=(-7/N_s^2+36/N_s^3)/F_{2,2}$, and $x_2=(8/N_s^2)/F_{2,2}$. The $F_{2,2}$ factor, equal to $1-10/N_s+33/N_s^2-36/N_s^3$, comes from the  norm, $N! N'! F_{N,N'}$, of the $\ket{\psi_{N,N'}}$ state given in Eq.~(\ref{FNN}). Fermion exchanges like the one of Fig.~\ref{fig:2}a give the $1/N_s$ term of $x_0$;  the one of Fig.~\ref{fig:2}b gives the $1/N^2_s$ term of $x_1$;  the one of Fig.~\ref{fig:2}c gives the $x_2$ prefactor of the $m=2$ term. Figure \ref{fig:3} shows the resulting correlation function for $N_s=5$, from which we clearly see the next-to-lowest oscillatory mode, and the singularity at $\bR_{12}=0$.

So far, we have considered bound fermion pair $a\dg_{\bR_i} b\dg_{\bR_i} $ with no relative motion extension. Pauli blocking then appears in the strongest way by forbidding two dimers to be at the same site through the $(1-\delta_{\bR_1,\bR_2})$ prefactor in Eq.~(\ref{eq:G2}). In reality, physical dimers are trapped in lattice potential wells with finite depth and width; so they have a finite spatial extension.  This physically broadens the effect of Pauli blocking and transforms the singular dip of Eq.~(\ref{eq:G2}) into a smooth dip (see Fig.~\ref{fig:3}). However, such dimer granularity must not affect the interference pattern at larger scale. Interestingly, a similar dip feature has been found for elementary bosons with hardcore repulsion\cite{Naraschewski1999}.


\section{Conclusion}  
In this work, we address the commonly bypassed consequences of the particle composite nature in cold-atom physics, by considering the interference pattern of two  condensates made of dimers. We predict the existence of additional high frequency modes, in contrast to a unique low-frequency mode existing when the particle composite nature is neglected. With the help of analytical calculations and Shiva diagrams that visualize composite boson many-body effects, we evidence that these high modes come from dimensionless fermion exchanges between condensates. Being many-body in essence, the amplitude of these high modes  depends on  density;  therefore, their observation requires rather dense condensates, that is, sizeable many-body parameter $\eta$, as possibly obtained by using optical lattices. In addition, Pauli blocking between the particle fermionic components produces a dip in the interference pattern that constitutes another signature of the dimer granularity.\

Just like the composite nature of semiconductor excitons has  revealed a breadth of remarkable effects, we anticipate cold-atom systems  to provide a novel, fully controllable playground to investigate further in depth the very unique many-body effects that result from dimensionless fermion exchanges, that is, exchange in the absence of fermion-fermion interaction. Recent  optical lattices already reach densities high enough for these new many-body effects to be observable, including  the signatures we here predict.


\ack
 
S.Y. S. and M. C. acknowledge support from OBELIX from the French Agency for Research (ANR-15-CE30-0020). A.C. thanks Geza Giedke for insightful comments. 
   
\appendix

\renewcommand{\thesection}{\mbox{Appendix~\Roman{section}}} 
\setcounter{section}{0}
\renewcommand{\theequation}{\mbox{A.\arabic{equation}}} 
\setcounter{equation}{0} %
\section{Elementary bosons\label{app:sec1}}
 
Elementary boson operators obey the commutation relations $[ \bar B\dg_{\bQ'}, \bar B\dg_{\bQ} ]_- = 0$ and $[ \bar B_{\bQ'}, \bar B\dg_{\bQ} ]_- = \delta_{\bQ'\bQ}$; so, by iteration, 
\begin{eqnarray}\label{eq:AN}
\left[ \bar B_{\bQ'},( \bar B\dg_{\bQ})^N \right]_-=N\,  \delta_{\bQ'\bQ} \, (\bar B\dg_\bQ)^{N-1}\,. \label{app:AAN_elm}
\end{eqnarray}
This commutator  immediately gives the normalization factor of the state $\ket{\bar \psi_{N,N'}} = (\bar B\dg_\bQ)^N  (\bar B\dg_{\bQ'})^{N'} \vac$ as ${ \braket{\bar\psi_{N,N'}}{\bar\psi_{N,N'}}  } =N!{N'}!$ where   $\vac$ denotes the vacuum, and with a little more work, the value of 
\begin{equation}
\mathcal{\bar  S}^{(n)}_{N,N'}(\{\bR_i\})=\bra{\bar\psi_{N,N'}} \bar B\dg_{\bR_1} \dots \bar B\dg_{\bR_n} \bar B_{\bR_n} \dots \bar B_{\bR _1} \ket{\bar\psi_{N,N'}  } \label{app:SN_elem}
\end{equation}
with $\bar B\dg_\bR = \sum_\bQ  \bar B\dg_\bQ\braket{\bQ}{\bR}$, that enters the $n$-particle spatial correlation function  
\begin{equation}\label{eq:Gn_elementary}
\mathcal{\bar  G}^{(n)}_{N,N'}(\{\bR_i\})=\frac{\mathcal{\bar{S}}^{(n)}_{N,N'} (\{\bR_i\}) }{\braket{\bar \psi_{N,N'}}{\bar \psi_{N,N'}}}\,.
\end{equation}
The $\bR$-space and $\bQ$-space creation operators are linked  through   $\braket{\bQ}{\bR} = e^{-i\bQ \cdot \bR} / L^{D/2}$ for continuous $\bR$ in a finite volume $L^D$, with  $L^D$ replaced by $N_s$  for a  discrete lattice  of $N_s$ sites located at $\{\bR_i\}$.\



In this work, we  propose an original procedure to evaluate the spatial correlation function (\ref{eq:Gn_elementary}). Although this procedure might appear as overcomplicated for elementary bosons,  it  allows handling more complex cobosons in an exact way. We introduce a generalized elementary-boson coherent state   
\begin{equation} 
\ket{\bar\phi_{z,z'}}=e^{z\bar B\dg_\bQ+z' \bar B\dg_{\bQ'}}\vac\, ,
\end{equation}
where $(z,z')$ are complex scalars. The residue theorem gives the two-condensate state $\ket{\bar\psi_{N,N'}}$ as  
\begin{equation}
\ket{\bar \psi_{N,N'}}=N!{N'}!\oint \frac{dz}{2\pi i}\frac{1}{z^{N+{N'}+1}} \oint \frac{dz'}{2\pi i}\frac{1}{ z'^{{N'}+1}}\ket{\bar \phi_{z,zz'}}.\label{app:ketNM}
\end{equation}
Turning from $\ket{\bar \phi_{z,z'}}$ to $\ket{\bar \phi_{z,zz'}}$   allows controlling  the {\it total} number $N+{N'}$ of bosons with momentum $\bQ$ and $\bQ'$ through $z$, and the number ${N'}$ of $\bQ'$ bosons  through $z'$. This will later  on facilitate expansion in the boson density through  $z$ factors.  

We note that  the only part of $\ket{\bar \phi_{1,1}}=e^{(\bar B\dg_\bQ+ \bar B\dg_{\bQ'})}\vac$  that gives a non-zero contribution when projected over $\ket{\bar{\psi}_{N,N'}}$ is the one that has the same particle number and momentum, i.e., $\ket{\bar \psi_{N,N'}}$ itself. This remark helps seeing that  
\begin{equation}
 (N!{N'}!)^2\oint \frac{dz}{2\pi i}\frac{1}{z^{N+{N'}+1}} \oint \frac{dz'}{2\pi i}\frac{1}{ z'^{{N'}+1}}\braket{\bar \phi_{1,1}}{\bar \phi_{z,zz'}}=\braket{\bar \psi_{N,N'}}{\bar \psi_{N,N'}}\,.\label{app:nor_elem}
\end{equation}
  In the same way, Eq.~(\ref{app:SN_elem})  can be rewritten as
\begin{eqnarray}
\mathcal{\bar S}^{(n)}_{N,N'} (\{\bR_i\})&=& (N!N'!)^2\oint \frac{dz}{2\pi i}\frac{1}{z^{N+N'+1}} \oint \frac{dz'}{2\pi i}\frac{1}{ z'^{N'+1}}\label{app:SnN_elm_01}  \nonumber \\
 &&\times \bra{\bar \phi_{1,1}}  \bar B\dg_{\bR_1} \dots \bar B\dg_{\bR_n} \bar B_{\bR_n} \dots \bar B_{\bR _1} \ket{\bar \phi_{z,zz'}}.\:\:\:
\end{eqnarray}

 The trick  now is to  calculate the above scalar product by using commutators in real space instead of momentum space  as in Eq.~(\ref{app:AAN_elm}). We first note that   $\ket{\bar \phi_{z,z'}}$ is eigenstate of the $\bar B_{\bR }$ operator,   
 \begin{equation} \label{eq:f_elem}
 \bar B_{\bR } \ket{\bar \phi_{z,z'}}= f_{z,z'}(\bR) \ket{\bar \phi_{z,z'}}\, ,
 \end{equation}
   with the eigenvalue $ f_{z,z'}(\bR)=z\braket{\bR}{\bQ}+z' \braket{\bR}{\bQ'}$. 
   So, we readily find 
\begin{equation}
\frac{\bra{\bar \phi_{1,1}}  \bar B\dg_{\bR_1} \dots \bar B\dg_{\bR_n} \bar B_{\bR_n} \dots \bar B_{\bR _1} \ket{\bar \phi_{z,z'}}}{ \braket{\bar \phi_{1,1}}{\bar \phi_{z,z'}}}=\prod_{i=1}^n G_{z,z'}(\bR_i)\equiv \frac{\bar \Xi^{(n)}_{z,z'}(\bR_i)}{L^{nD}} \, ,\label{app:BB11zz'_elem}
\end{equation} 
with $G_{z,z'}(\bR)= f^*_{1,1}(\bR) f_{z,z'}(\bR) $. The major  advantage of this new procedure is to avoid enforcing  momentum conservation at each commutation step; instead, the relevant momentum-conserving processes are selected at the very end only, directly through  $\bar \Xi^{(n)}_{z,z'}(\{\bR_i\})$: indeed, for $n=1$, Eq.~(\ref{app:BB11zz'_elem}) readily gives $\Xi^{(1)}_{z,z'}(\bR)=z+z'+z e^{i\bR\cdot(\bQ-\bQ')}+z' e^{i\bR\cdot(\bQ'-\bQ)}$, in which the terms that conserve momentum are $z+z'$, so that  $\Xi^{(1)}_{z,z'}(\bR)$ must reduce to  $ z+z'$. In the same way,  $ \bar \Xi_{z,z'}^{(n)}(\{\bR_i\})$ for $n=2$ is equal  to $ (z+z')^2+2zz' \cos(\bQ-\bQ')\cdot \bR_{12}$; for $n=3$  it is equal to $(z+z')^3+2zz'(z+z') \big(\cos(\bQ- \bQ') \cdot \bR_{12}+\cos(\bQ- \bQ') \cdot \bR_{23}+\cos(\bQ- \bQ') \cdot \bR_{31}\big)$  with $\bR_{ij}=\bR_i-\bR_j$, and so on...  The above results used  for the scalar product in Eq.~(\ref{app:SnN_elm_01}) give, with the help of  Eq.~(\ref{app:nor_elem}), the first $n$-particle correlation functions for free elementary bosons as 
\begin{eqnarray}
\hspace{-2.5cm}\mathcal{\bar  G}^{(1)}_{N,N'}(\bR_1)=\frac{1}{L^D}(N+N')\,,\\
\hspace{-2.5cm} \mathcal{\bar  G}_{N,N'}^{(2)}(\bR_1,\bR_2){=}\frac{1}{L^{2D}} \Big\{N(N-1)+N'(N'-1)+2NN'+ 2 NN' \cos (\bQ- \bQ') {\cdot} \bR_{12} \Big\}\,,\\
\hspace{-2.5cm}\mathcal{\bar  G}^{(3)}_{N,N'}(\bR_1,\bR_2,\bR_3)=\frac{1}{L^{3D}}  \Big \{N(N-1)(N-2)+ N'(N'-1)(N'-2) + 3 N' N (N -1) \nonumber \\
+ 3 N N' (N'-1)  + \big(2 N' N (N-1)+2 N N' (N'-1)\big) \Big(\cos (\bQ-\bQ') {\cdot} \bR_{12} \nonumber \\
 + \cos (\bQ-\bQ') {\cdot} \bR_{23}+ \cos (\bQ-\bQ') {\cdot} \bR_{31}\Big)\Big\} \,,
\end{eqnarray} 
with $L^D$ replaced by $N_s$ in the case of  discrete $\bR_i$'s.\

The 1-particle function $\mathcal{\bar  G}^{(1)}_{N,N'}(\bR_1)$ physically corresponds to the total  density of the  $\bQ$ and $\bQ'$  elementary bosons in the sample volume $L^D$, while the other results evidence that the collision of two elementary-boson condensates  leads to wave-like interference patterns associated with the momentum difference $(\bQ - \bQ')$. This pattern can be observed by measuring the $n$-particle correlation function $\mathcal{\bar  G}^{(n)}_{N,N'}(\{\bR_i\})$ for $n\geq 2$. The expression of this correlation function  for $N=N'$ and $\bQ'=-\bQ$ has already been found in \cite{Javanainen1996a}.\

 It can be of interest to note that the quantity 
\begin{eqnarray} \label{eq:Pbosons}
\bar P_N^{(n)}(\{\bR_i\}) = L^{nD}  \frac{(2N-n)!}{(2N)!} \mathcal{\bar  G}^{(n)}_{N,N}(\{\bR_i\})\,,
\end{eqnarray}
corresponds to the probability of detecting $n$ bosons located at $(\bR_1, \dots, \bR_n)$ in the $\ket{\bar \psi_{N,N}}$ condensate, as suggested in \cite{Javanainen1996a}: Indeed, $\bar P^{(1)}_N(\bR_1)=1$ while the $\bar P^{(n)}_N$'s are linked by 
\begin{equation}\label{5}
\int \frac{d\bR_n}{L^D}\bar  P^{(n)}_N(\bR_1, \dots, \bR_n) = \bar P^{(n-1)}_N (\bR_1, \dots, \bR_{n-1}),
\end{equation}
as physically required for probabilities. In the case of composite bosons, the correlation functions  have  additional terms induced by the fermion exchanges that prevent such  identification.

 \renewcommand{\theequation}{\mbox{B.\arabic{equation}}} 
\setcounter{equation}{0}
\section{Composite bosons: fermionic-atom dimers\label{app:sec2}}
%


We now consider an optical lattice of $N_s$ sites, each site possibly hosting a bound pair of
different fermionic atoms, with creation operator $\fB\dg_{\bR_i}=a\dg_{\bR_i} b\dg_{\bR_i}$. Due to inter-site interaction, the resulting coboson dimer creation operators read $ \fB\dg_{\bQ} = \sum_{i=1}^{N_s} \fB\dg_{\bR_i} \braket{\bR_i}{\bQ}$, with  $\braket{\bR_i}{\bQ} = e^{i\bQ\cdot \bR_i} /\sqrt{N_s}$. They obey the commutation relations 
\begin{eqnarray}
\left[\fB_{\bQ'}, \fB\dg_\bQ\right]_- &=& \delta_{\bQ' \bQ} - \fD_{\bQ' \bQ}\, ,  \label{BBcommu}\\
 \fD_{\bQ' \bQ} &=& \frac{1}{N_s}\sum_{j=1}^{N_s} e^{-i(\bQ'-\bQ)\cdot\bR_j}  \left(a\dg_{\bR_j} a_{\bR_j} + b\dg_{\bR_j} b_{\bR_j}\right) .\:  \label{Dfunc}
\end{eqnarray}         
 As usual\cite{MoniqueSeanbook}, the deviation-from-boson operator $\fD_{\bQ' \bQ}$ generates the dimensionless Pauli scatterings $\lambda \left(_{\bQ'_1 \bQ_1}^{\bQ'_2 \bQ_2}\right)$ responsible for  fermion exchanges  between  cobosons. In the case of the single-index cobosons we here consider, they reduce to 
 \begin{eqnarray} \label{eq:com_Frenkel}
\left[ \fD_{\bQ_1' \bQ_1},  \fB\dg_{\bQ_2} \right]_- &=& \sum_{\bQ'_2} \fB\dg_{\bQ'_2} \left\{ {\lambda} \left(_{\bQ'_1 \bQ_1}^{\bQ'_2 \bQ_2}\right) + (\bQ_1 \longleftrightarrow \bQ_2) \right\} \nonumber\\
&=&\frac{2}{N_s} \fB\dg_{\bQ_1 + \bQ_2 - \bQ'_1} \, .\label{DBcommu}
\end{eqnarray}

The  correlation function for detecting $n$ dimers for the state $\ket{\psi_{N,N'}} = (\fB\dg_\bQ)^N  (\fB\dg_{\bQ'})^{N'} \vac$ reads as $\mathcal{G}_{N,N'}^{(n)}(\{\bR_i\})=\mathcal{S}_{N,N'}^{(n)}(\{\bR_i\})/\braket{\psi_{N,N'}}{\psi_{N,N'}}$ with
\begin{equation}
\mathcal{S}_{N,N'}^{(n)}(\{\bR_i\}) = \bra{\psi_{N,N'}} \fB\dg_{\bR_1} \dots \fB\dg_{\bR_n}  \fB_{\bR_n} \dots \fB_{\bR_1} \ket{\psi_{N,N'}}\,.\label{app:SNncobo_00}
\end{equation}
Here also, we introduce the  generalized composite-boson coherent state $\ket{\phi_{z,z'}}=e^{z\fB\dg_\bQ+z' \fB\dg_{\bQ'}}\vac$. As for elementary bosons, we  can rewrite Eq.~(\ref{app:SNncobo_00}) as
\begin{eqnarray}
 \mathcal{S}^{(n)}_{N,N'} (\{\bR_i\})&=& (N!N'!)^2\oint \frac{dz}{2\pi i}\frac{1}{z^{N+N'+1}} \oint \frac{dz'}{2\pi i}\frac{1}{ z'^{N'+1}} \nonumber \\
 &&\times \bra{ \phi_{1,1}}  \fB\dg_{\bR_1} \dots \fB\dg_{\bR_n}  \fB_{\bR_n} \dots \fB_{\bR_1}  \ket{ \phi_{z,zz'}}\,. 
\end{eqnarray}
The procedure is essentially the same as   for  elementary bosons, equations (\ref{app:ketNM},\ref{app:nor_elem},\ref{app:SnN_elm_01}) staying  valid for cobosons. 
 Momentum conservation at each commutation is even more cumbersome due to additional fermion exchanges, which are many-body in nature. This is why working with commutators in real space is really advantageous.

It will appear as convenient to first note that the deviation-from-boson operator $\fD_{\bR \bQ}=\sum_\bP \braket{\bR}{\bP} \fD_{\bP\bQ}$ leads to
\begin{equation}
\left[ \fD_{\bR \bQ}, e^{z\fB\dg_{\bQ'}} \right]_- = 2z\braket{\bR}{\bQ}\braket{\bR}{\bQ'} \fB\dg_{\bR}e^{z\fB\dg_{\bQ'}}. \label{app:fDbRbQ}
\end{equation}
 In the same way, 
\begin{equation} \label{app:fBTzQzQ}
\left[\fB_{\bR}, e^{z\fB\dg_\bQ+z'\fB\dg_{\bQ'}}\right]_- =  e^{z\fB\dg_\bQ+z'\fB\dg_{\bQ'}} \Big(\mathcal{F}^+_{z,z'}(\bR)- z\fD_{\bR \bQ}-z'\fD_{\bR \bQ'}\Big)\, , 
\end{equation}
with 
\begin{equation}\label{F+}
{\mathcal{F}}^+_{z,z'}(\bR)=f_{z,z'}(\bR) \left \{1 - f_{z,z'}(\bR)\fB\dg_{\bR}\right \}=f_{z,z'}(\bR) e^{-  f_{z,z'}(\bR) \fB\dg_\bR}\,,
\end{equation} 
 since $(\fB\dg_\bR)^2=0$. 

The above two commutators are obtained from iteration of the coboson commutation relations in momentum space, Eqs.~(\ref{BBcommu},\ref{DBcommu}), namely
\begin{eqnarray} \label{eq:BN_Frenkel}
\hspace{-1cm}\left[\fB_{\bQ'}, (\fB\dg_\bQ)^N\right]_- = N (\fB\dg_\bQ)^{N-1} (\delta_{\bQ' \bQ} - \fD_{\bQ' \bQ}) - \frac{N(N-1)}{N_s} (\fB\dg_\bQ)^{N-2}  \fB\dg_{2\bQ-\bQ'}\, ,  \\
\hspace{-1cm}\left[ \fD_{\bQ'_1 \bQ_1},  (\fB\dg_{\bQ})^N \right]_- = 2\frac{N}{N_s} (\fB\dg_{\bQ})^{N-1}  \fB\dg_{\bQ + \bQ_1 - \bQ_1'}.
\end{eqnarray}
Equation (\ref{app:fBTzQzQ}) then gives
 \begin{equation}
\fB_{\bR}\ket{\phi_{z,z'}} = \mathcal{F}^+_{z,z'}(\bR)\ket{\phi_{z,z'}}. \label{app:fBR1ketz1z2}
\end{equation}

The  curly bracket  in Eq.~(\ref{F+}),  absent for elementary bosons (see Eq.~(\ref{eq:f_elem})), results from fermion exchanges occurring within  the $\ket{\phi_{z,z'}}$  state. It makes $\ket{\phi_{z,z'}}$  not an eigenstate of the fermion pair operator $\fB_{\bR}$.  In the same way, we find 
\begin{equation}
\fB_{\bR_2}\fB_{\bR_1}\ket{\phi_{z,z'}} =\big \{ 1-\delta_{\bR_2\bR_1}\big\} {\mathcal{F}}^+_{z,z'}(\bR_2) {\mathcal{F}}^+_{z,z'}(\bR_1)\ket{\phi_{z,z'}}\, ,\label{app:fBR1R2ketz1z2}
\end{equation}
 the curly bracket  coming from  Pauli blocking as  $\fB_{\bR_2}\fB_{\bR_1}=0$ when $\bR_2=\bR_1$; and so on...\


The major advantage of using commutators in real space is that $(\fB\dg_{\bR_i})^n$ readily gives zero  for $n\geq 2$ whenever it appears. Equations (\ref{F+},\ref{app:fBR1ketz1z2}) then give
\begin{equation}
\bra{\phi_{1,1}} \fB\dg_{\bR} \fB_{\bR} \ket{\phi_{z,z'}} = f_{z,z'}(\bR) \bra{\phi_{1,1}} \fB\dg_{\bR}  \ket{\phi_{z,z'}} \,,\label{app:fBR1ketz1z2_01}
\end{equation}
while   for $\bR_1\not= \bR_2$, Eq.~(\ref{app:fBR1R2ketz1z2}) gives
\begin{equation}
\bra{\phi_{1,1}} \fB\dg_{\bR_1}  \fB\dg_{\bR_2}  \fB_{\bR_2} \fB_{\bR_1} \ket{\phi_{z,z'}}=   f_{z,z'}(\bR_1) f_{z,z'}(\bR_2) \bra{\phi_{1,1}} \fB\dg_{\bR_1}  \fB\dg_{\bR_2}  \ket{\phi_{z,z'}}\, .
\end{equation}
More generally, $\bra{\phi_{1,1}} \fB\dg_{\bR_1}\dots  \fB\dg_{\bR_n}  \fB_{\bR_n}\dots \fB_{\bR_1} \ket{\phi_{z,z'}}$ reduces for different $\bR_i$'s  to
\begin{equation}
 \left(\prod_{i=1}^n f_{z,z'}(\bR_i)\right) \bra{\phi_{1,1}} \fB\dg_{\bR_1}\dots  \fB\dg_{\bR_n} \ket{\phi_{z,z'}} \,.
\end{equation}

If we now use Eq.~(\ref{app:fBR1ketz1z2}) for $\bra{\phi_{1,1}}\fB\dg_\bR$, we end with
\begin{eqnarray}
\frac{\bra{\phi_{1,1}} \fB\dg_{\bR} \fB_{\bR} \ket{\phi_{z,z'}}}{\braket{\phi_{1,1}}{\phi_{z,z'}}} =\frac{G_{z,z'}(\bR)}{1+G_{z,z'}(\bR)} \,,\label{app:fBR1ketz1z2_02}
\end{eqnarray} 
and more generally,
\begin{equation}
\frac{\bra{\phi_{1,1}} \fB\dg_{\bR_1}\dots  \fB\dg_{\bR_n}  \fB_{\bR_n}\dots \fB_{\bR_1} \ket{\phi_{z,z'}}}{\braket{\phi_{1,1}}{\phi_{z,z'}}}=\prod_{i=1}^n\frac{G_{z,z'}(\bR_i)}{1+G_{z,z'}(\bR_i)}\, .
\label{app:fBR1R2ketz1z2_02}
\end{equation}
This result can also be obtained from the definition of the state $\ket{\phi_{z,z'}}$ and the relation $\fB\dg_{\bQ} = \sum_{\bR} \fB\dg_{\bR} \braket{\bR}{\bQ} $, so that
\begin{eqnarray}
\ket{\phi_{z,z'}} &=& e^{\sum_\bR (z \braket{\bR}{\bQ} + z' \braket{\bR}{\bQ'}) \fB\dg_{\bR}} \vac \nonumber \\
&=& \prod_\bR \left(1+ f_{z,z'}(\bR) \fB\dg_{\bR} \right) \vac.
\end{eqnarray}
Comparing the result for coboson dimers (\ref{app:fBR1R2ketz1z2_02}) with that for elementary bosons (\ref{app:BB11zz'_elem}), we can trace the denominator in the RHS of Eq.~(\ref{app:fBR1R2ketz1z2_02}) back to the curly bracket of $\mathcal{F}^+_{z,z'}(\bR)$ given in Eq.~(\ref{F+}).

%
%

Let us now focus  on the  correlation functions for one and two dimers. Extension to multiple dimers is straightforward. As for elementary bosons,  the relevant momentum-conserving processes  are selected from 
\begin{equation}
\Xi^{(n)}_{z,zz'}(\{\bR_i\})= \prod_{i=1}^n\frac{G_{z,zz'}(\bR_i)}{1+G_{z,zz'}(\bR_i)}.
\end{equation} 
We first expand in $z$, which is easy to do by  noting that $f_{z,{zz'}}(\bR)=z  f_{1,{z'}}(\bR)$.   For $n=1$, this gives 
\begin{equation} \label{expansion}
 \sum_{p=1}^{\infty} (-1)^{p-1} \left(\frac{z}{N_s} \right)^{p} \left(e^{-i \bR \cdot \bQ} + e^{-i \bR \cdot \bQ'}\right)^{p}\left(e^{i \bR \cdot \bQ} + z' e^{i \bR \cdot \bQ'}\right)^{p} .
\end{equation}
Selecting  momentum-conserving processes yields 
\begin{equation} \label{app:fBR1ketz1z2_03}
\Xi^{(1)}_{z,zz'}(\bR) = \sum_{p=1}^\infty (-1)^{p-1} \! \left(\frac{z}{N_s}\right)^p  \sum_{m=0}^p  (C^p_m)^2  (z')^m\, ,
\end{equation}
where $C^p_m$ denotes the binomial coefficient $\binom{p}{m}$. 
Similarly, for $n=2$, we find 
\begin{eqnarray}
\hspace{-1cm}\Xi^{(2)}_{z,zz'}(\bR_1, \bR_2)  = \sum_{p=1}^\infty \sum_{p'=1}^\infty \left( \frac{-z}{N_s} \right)^{p + p'}\sum_{m_1=0}^{p} \sum_{m_1'=0}^{p} e^{(m'_1 -m_1)i \bR_{12} \cdot (\bQ - \bQ')} C^{p}_{m_1} C^{p}_{m'_1}  \\
\times\sum_{m_2=0}^{p'} (z')^{m_1+m_2} C^{p'}_{m_2} C^{p'}_{ m_1 +m_2-m'_1}\,. \nonumber
\end{eqnarray}

To go further and obtain the spatial correlation function, we need the normalization factor $\braket{\psi_{N,N'}}{\psi_{N,N'}}$. This quantity  is quite tricky to derive from a naive expansion, because fermion exchange  not only occurs between dimers carrying same momentum but also between dimers carrying different momenta. The same procedure, that is,   Eq.~(\ref{app:nor_elem}) rewritten for dimers, gives 
\begin{eqnarray}
F_{N,{N'}}&=&\frac{\braket{\psi_{N,N'}}{\psi_{N,N'}}}{N!{N'}!} \label{app:nor_elem_02}\\
&=&N!{N'}!\oint \frac{dz}{2\pi i}\frac{1}{z^{N+{N'}+1}} \oint \frac{dz'}{2\pi i}\frac{1}{ z'^{{N'}+1}}\braket{\phi_{1,1}}{\phi_{z,zz'}},\nonumber
\end{eqnarray}
which also reads, through an integration by part over $z$, as
\begin{equation}
F_{N,{N'}}=\frac{N!{N'}!}{N+{N'}}\oint \frac{dz}{2\pi i}\frac{1}{z^{N+{N'}}} \oint \frac{dz'}{2\pi i}\frac{1}{ z'^{{N'}+1}} \bra{\phi_{1,1}} \fB\dg_{\bQ}+z'\fB\dg_{\bQ'} \ket{\phi_{z,zz'}}.\label{app:cobons_FM_02}
\end{equation}
This quantity is best calculated from  commutators in real space through  $\fB\dg_{\bQ}=\sum_\bR \braket{\bR}{\bQ} B\dg_\bR$. 
We then find  
\begin{eqnarray}
\frac{\bra{\phi_{1,1}} \fB\dg_{\bQ}+z'\fB\dg_{\bQ'} \ket{\phi_{z,zz'}}}{\braket{\phi_{1,1}}{\phi_{z,zz'}}}= \sum_\bR\frac{G_{1,z'}(\bR)}{1+G_{z,zz'}(\bR)}, 
\end{eqnarray}
as obtained by using Eq.~(\ref{app:fBR1ketz1z2}).
The above sum over $\bR$ has the effect of selecting  momentum-conserving processes, as  obtained through an $z$ expansion similar to the one performed in Eq.~(\ref{expansion}).   Equation (\ref{app:cobons_FM_02}) then leads to
\begin{equation}\label{FNN}
F_{N,N'} = \frac{1}{N+N'} \sum_{p=1}^{N+N'}p!\left( \frac{-1}{N_s}\right)^{p-1} \sum_{m={\rm max}\{p-N,0\}}^{{\rm min}\{p,N'\}}C^N_{p-m}C^{N'}_m C^p_m \: F_{N+m-p,N'-m}\,.
\end{equation}
This equation provides an efficient iteration to obtain high $F_{N,{N'}}$ terms, starting from $F_{0,0}=1$. The first ones read 
\begin{subequations}
\begin{align}
    & F_{1,1}= 1 -\frac{2}{N_s}, \label{eq:subeq1} \\
     F_{0,2} = ~&F_{2,0} = 1 -\frac{1}{N_s}, \label{eq:subeq2} \\
     F_{1,2}=~ & F_{2,1} = 1 -\frac{5}{N_s} +\frac{6}{N_s^2}, \label{eq:subeq3}\\
    & F_{2,2} =  1 -\frac{10}{N_s} +\frac{33}{N_s^2} - \frac{36}{N_s^3}. 
\end{align}
\end{subequations}

 The 1-dimer function   follows from Eq.~(\ref{app:fBR1ketz1z2_03}) divided by the norm of the $\ket{\psi_{N,N'}}$ state, as obtained  from Eq.~(\ref{FNN}). We find that it  simply reduces to the dimer density, namely $\mathcal{G}_{N,N'}^{(1)}(\bR)=(N+N')/N_s$, as obtained for elementary bosons. The physical reason for  not having any correction is that it fundamentally deals with detecting a single dimer. \

 The 2-dimer correlation function is modified by terms stemming  from fermion exchanges. For $n=2$, the result already is quite complicated,
\begin{eqnarray}\label{app:generalGNN}
\hspace{-2cm}\mathcal{G}_{N,N'}^{(2)}(\bR_1,\bR_2)=\sum_{p=1}^{N+N'-1}\sum_{p'=1}^{N+N'-p}\frac{(-1)^{p+p'}}{N_s^{p+p'}}\sum_{m_1=0}^{p}\sum_{m_1'=0}^p C^{p}_{m_1} C^{p}_{m'_1}  e^{(m'_1-m_1)i\bR_{12}\cdot(\bQ-\bQ')}\label{app:genraGN22}\\
\times \sum_{m_2={\rm max}\{0,m'_1-m_1\}}^{{\rm min}\{p',N'-m_1\}}C^{p'}_{m_2}C^{p'}_{m_1+m_2-m'_1} (p+p'-m_1-m_2)!  \nonumber \\
\times C^N_{p+p'-m_1-m_2} (m_1+m_2)!C^{N'}_{m_1+m_2}\frac{F_{N-p-p'+m_1+m_2,N'-m_1-m_2}}{F_{N,N'}}\, .\nonumber
\end{eqnarray}
When $N=N'=1$, it reduces to 
\begin{equation}
\mathcal{G}_{1,1}^{(2)}(\bR_1,\bR_2) = \frac{2}{N_s^2 F_{1,1}}\big\{1+  \cos (\bQ-\bQ')\cdot\bR_{12}\big\}\,.
\end{equation}
From the general form (\ref{app:genraGN22}),  we see that interferences in $\cos |m_1-m'_1|(\bQ-\bQ')\cdot\bR_{12}$ with $|m_1-m'_1|\not=1$ do exist for $\{N,N'\}\ge \{2,2\}$.
The explicit result for $N=N'=2$ is given in Eq.~(\ref{G2_2}).

\end{document}